\UseRawInputEncoding
\documentclass[conference]{IEEEtran}
\IEEEoverridecommandlockouts
\usepackage{cite}
\usepackage{amsmath,amssymb,amsfonts}
\usepackage{algorithmic}
\usepackage{graphicx}
\usepackage{dblfloatfix}
\usepackage{textcomp}
\usepackage{xcolor}
\def\BibTeX{{\rm B\kern-.05em{\sc i\kern-.025em b}\kern-.08em
    T\kern-.1667em\lower.7ex\hbox{E}\kern-.125emX}}

\begin{document}

\title{{\footnotesize 2020 Conference on Self-Organising Systems at Christian-Albrechts University in Kiel}\\Combining SimTrust and Weighted Simple Exponential Smoothing}

\author{\IEEEauthorblockN{Tobias Michel Latta}
\IEEEauthorblockA{\textit{Intelligent Systems} \\
\textit{Christian-Albrechts University}\\
Kiel, Germany \\
stu224916@mail.uni-kiel.de}
}

\maketitle
\begin{abstract}
In the domain of Autonomic and Organic Computing, the entities of a distributed system are variable as well as the efficiency and the intention of their work. Therefore, a scalable mechanism to incentivise/sanction entities which contribute towards/against the system goal is needed. Trust is a suited metric to find benevolent entities. In this paper, we focus for one on the SimTrust model which introduces trust on entities when they share interest and opinions using tagging information. The second model is the Weighted Simple Exponential Smoothing Trust metric (WSES) which functions on explicitly rated items. WSES follows two basic rules which ensure a logic rating mechanism. When putting these two models in context, SimTrust has advantages on items that have not been rated yet or can not easily be rated. WSES is a trust metric which returns good results on explicit rank values. We propose concepts on combining both approaches and state in which cases they are incompatible.
\end{abstract}


\section{Introduction}

\textmd{In every open system, where multiple entities interact with each other, trust plays a meaningful part. In self-organising systems (SOS), a successful interaction between two entities may only be performed, when both parties can estimate the behaviour of one another. Because these systems are getting more accessible, the number of interactable entities is growing. With that, the entities are increasingly unfamiliar with each other \cite{Tomforde2014}. To establish successful cooperations, trust metrics are integrated in the architecture of the SOS. This is often done via Collaborative Filtering (CF). CF takes into account the behaviour of every entity in the SOS. Similar entities get clustered and recommendations are created based on the behaviour of the other entities in the cluster \cite{Schafer2007},\cite{Massa2007}.\\
This implies that with increasing user activity the quality of clustering and recommendations is improving. On the other hand, for little interaction or novel entities, the quality of recommendations is not sufficient. This is typically referred to as the cold-start problem \cite{Armada2011}. Focussing on this issue is SimTrust where the entity has to provide personalized tags upon registration. This way, no explicit rating is needed to find quantified similarities on entities \cite{Simtrust2010}.\\
Besides the chosen fundamentals of trust derivation (e.g. CF or SimTrust), it is also important which metric is used. An intuitive metric might be rating all interactions on a certain scale. The Weighted Simple Exponential Smoothing metric (WSES) takes into account a set of rules which need to be followed to ensure an intuitive and fair metric \cite{Kantert2015}. However, there are many different approaches to this and the question arises which approach is best-suited for a typical SOS.\\
The contribution of this paper is comparing the performance of SimTrust and WSES on certain use cases \cite{Sterlin2017}, \cite{Schafer2000}, \cite{Burke2011} and defining possible combinations where they can or can not benefit from each other.\\
In Section 2, the use of trust in SOS is explained with a focus on deriving from self-tagging by entities and explicit rating of thirds. In Section 3, trust from tagging is introduced in connection to the implementation via SimTrust. Afterwards this is done analogous for trust from rating with the WSES metric in Section 4. After both approaches for establishing trust have been described, they are compared in Section 5. SimTrust and WSES are applied in theory on different use cases as well as combined to see possible beneficial outcomes. In Section 6, a conclusion as well as a final discussion is provided.}

\begin{figure*}[bp]
	\centering
	\includegraphics[page=1, width=0.90\textwidth]{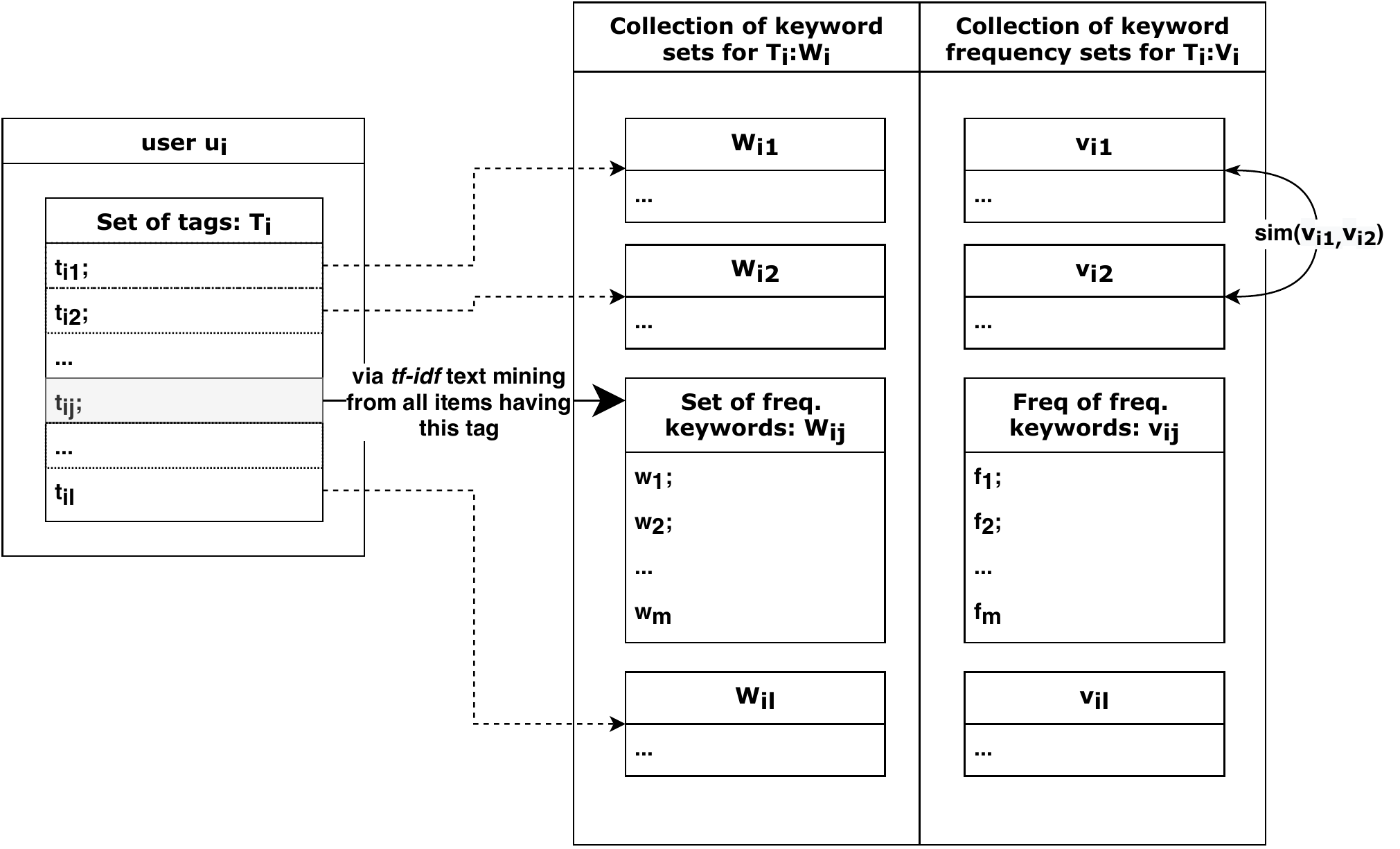}
	\caption{Visualization of the user-tag-keyword system}
	\label{fig_1}
\end{figure*}

\section{Trust in Self-organising systems}

\textmd{Self-organising systems are characterized by the varying participation and behaviour of entities. Due to missing control and surveillance over participants, a decentralized concept is needed that stimulates the contribution of entities towards a common system goal. In particular, the entities have to take over the responsibility to find and interact with suitable others as they are completely autonomous and may belong to different authorities. They have to find cooperation partners that reach a certain amount of efficiency and share the intention of being benevolent towards a common system goal \cite{Anders2016}.\\
To support this process, malevolent or inefficient entities have to be isolated. This can be done via computational trust which serves as a framework for entities to select and interact with others \cite{Anders2016}.\\
A standard approach to quantify trust for the system and the entities is CF. Similarities in previous activities and the known characteristics of two entities imply trust among them \cite{Bhuiyan2013}. This approach is for example applied on item recommendation in e-commerce \cite{Rayport2003}, \cite{Linden2003}.\\
Besides this technique, there are many more ways of identifying trust. By providing tags the entities must enter information, as it is done for content-based filtering \cite{Armada2011}. This information can be used to group similar entities which implies trust among them, as it will be explained in the following section.}

\section{Trust from Tagging - SimTrust}

\textmd{The first approach on trust addressed in this paper is presented in this section. To build a basic understanding trust derived from tagging is introduced in the first subsection. Building on that, the implementation via SimTrust will be covered in the second subsection.}

\subsection{Tagging as a Trust indicator}

\textmd{When in content-based filtering the user of an e-commerce platform has selected an item with a certain set of characteristics, it is likely to get another item recommended that has similar characteristics. So when entities have to define features for themselves, it is possible to group them by these. This can be expressed via multiple short labels or keywords, called tags \cite{Simtrust2013}. Research shows that there is a correlation between similar entity-behaviour and trust \cite{Ziegler2007}. In the context of SOS, this can be helpful for entities to solve problems in cooperation \cite{Bhuiyan2013}.}

\subsection{Implementation via SimTrust}

\begin{figure}[tbp]
		\centering
		\includegraphics[page=1, scale=0.7]{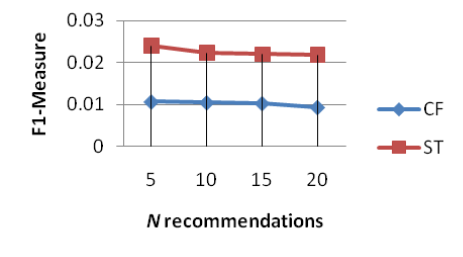}
		\caption{Recommender evaluation of SimTrust (ST) and Collaborative Filtering (CF) by Bhuiyan et al.}
		\label{fig_2}
	\end{figure}

	\textmd{SimTrust is an approach by Bhuiyan et al. \cite{Simtrust2010}. It uses trust for recommending items to entities based on other entities, they are trusting. When they are similar, they get recommended preferred or already acquired items from trusting entities. These similarities are quantified in the form of shared tags. Because the item tags will be set by the entities, they lack in standardization and unification. Therefore, the first step of SimTrust is to derive the semantic meaning of a tag by connecting it with keywords derived from the item description. In the following Fig.~\ref{fig_1}, the connections between entities (users in this case), tags and keywords are presented. A user $u_{i}$ (referenced by $i$) defines himself with a set of tags $T_{i}$. The semantics of the tags are derived from a text-mining procedure called \textit{tf-idf} \cite{Uther2010}. It connects a tag $t_{ij}$ with a set of keywords $w_{ij}$ and their frequencies $v_{ij}$. The $i$ references the user while the $j$ references the specific tag $j$. The frequency of a keyword is a measure for the importance of defining the tag. Tags $t_{ij}$ and $t_{ik}$ (another tag from the same user) are considered \textit{similar} when a function $sim(v_{ij},v_{ik})$ reaches a certain threshold.\\
	The question arises how it is possible to calculate a trust value between two users $u_{i}$ and $u_{j}$. This is solved by calculating the similarity of interest for every keyword $k$. These values are summed and averaged to express the overall similarity of $u_{i}$ and $u_{j}$.}
	\textmd{This concept has been tested on a dataset with 2,200 users and 18,663 books in comparison to a classic CF algorithm that would quantify trust among users on the overlap of previous ratings using Jaccard's coefficient \cite{Tan2017}. To compare the algorithms, all users were assigned with personal lists of preferred items $T_{i}$. Later, the recommendations made by the algorithms $P_{i}$ were compared to $T_{i}$ via a popular method called "{}Precision and Recall"{} by Cleverdon \cite{Cleverdon1966} in combination with a performance metric called F1 by Sarwar et al. \cite{Sarwar2002}.\\
	Fig.~\ref{fig_2} from Bhuiyan et al. \cite{Simtrust2010} shows an improvement by 200\% when comparing the performance of traditional collaborative filtering (CF) to SimTrust (ST). 
	}

\section{Trust from Rating - WSES}

\textmd{The second approach on trust is presented in this section. After the basics for Trust derived from ratings are explained, a specific implementation via WSES will be covered in the second subsection.}

\subsection{Rating as a Trust indicator}
	
	\textmd{In comparison to the previous tag-based approach, using explicit ratings seems to be a more intuitive way in establishing trust between entities. From school grades, IMDB ratings or sport disciplines: linear number-based ratings, like variants of the Likert scale \cite{Joshi2015}, aim for explicitness and comprehensibility even for novel participants of a system. Because of the quantifiability, ratings serve as a suited base for CF. The algorithm can easily group entities with similar ratings on similar items\cite{Herlocker2002}.\\
	On the other hand, ratings are very subjective due to the different requirements entities are setting for the items \cite{Ruan2016}, \cite{Bhuiyan2013}. Another issue is the rating metric itself. The value ranges are typically binary or float values. But how the personal ratings of an entity are calculated to an overall rating of an item is not as intuitive as it seems. Therefore, Kantert et al. \cite{Kantert2015} have developed a rating metric that fulfills two requirements which ensure a fair and realistic rating value for an item or entity. This metric is presented in the following subsection.}

\subsection{Implementation via Weighted SES Metric}

	\textmd{In the context of the previously mentioned Autonomic Computing \cite{Kephart2003} and Organic Computing \cite{MuellerTomforde2017} initiatives, trust has been considered as a means to assess the expected behaviour of participants in open, self-organising system constellations. Kantert et al. \cite{Kantert2015} have developed a metric for accumulating explicit ratings which prevents two major problems that intuitive metrics have. For example is normalizing the sum of ratings faulty in terms of incorporating the number of ratings, an entity received. The following example by Kantert et al. \cite{Kantert2015} illustrates this behaviour:}
	
	\begin{equation}\label{normsum1}
		R_{A} := (1,1,1)	
	\end{equation}
	\begin{equation}\label{normsum2}
		F(R_{A}) := \frac{1+1+1}{3} = 1	
	\end{equation}
	\begin{equation}\label{normsum3}
		R_{B} := (1,1,1,0.5,0.5,0.5)	
	\end{equation}
	\begin{equation}\label{normsum4}
		F(R_{B}) := \frac{3*1+3*0.5}{6} = 0.75	
	\end{equation}
	
	\textmd{In this exemplary SOS, entities A and B perform tasks for which they receive a real number rating between 0 and 1 ($\mathfrak{R}_{+}$), depending on the complexity or length of the task. Entity A has completed three tasks that are rewarded with three good ratings (\ref{normsum1}). The normalized average value is therefore also a good value (\ref{normsum2}). Another entity B has done a doubled amount of operations with three good ratings and three medium ratings (\ref{normsum3}). Intuitively, B should have a better overall rating than A because it fulfilled twice as many operations as A did and also received three good ratings. On the contrary, the normalized average value is lower than for A (\ref{normsum4}).\\
	From this, two requirements are derived: R1 describes the behaviour of a metric when two ratings $r_{1}$ and $r_{2}$ are added to a set of ratings $R_{n}$. For adding, the function $\mathfrak{U}$ is introduced. $r_{1}$ is bigger than $r_{2}$. The final reputation $\mathfrak{T}(R_{n})$ with $r_{1}$ has to be higher than $\mathfrak{T}(R_{n})$ with $r_{2}$ unless $R_{n}$ with $r_{2}$ is already the maximum value. The second requirement R2 ensures that every positive rating increases the reputation until the maximum value is reached. This prevents the faulty behaviour of the initial example.}

	\begin{equation}
		\begin{gathered}
			\forall r_{1}, r_{2} \in \mathfrak{R}_{+} : \mathfrak{T}(\mathfrak{U}(R_{n},r_{1})) > \mathfrak{T}(\mathfrak{U}(R_{n},r_{2}))	\vee \\
			\mathfrak{T}(\mathfrak{U}(R_{n},r_{2})) = 1, \\
			r_{1} > r_{2} \Rightarrow |\mathfrak{R}_{+}| > 1
		\end{gathered}
		\tag{R1}
	\end{equation}
	\begin{equation}
		\forall r \in \mathfrak{R}_{+} : \mathfrak{T}(\mathfrak{U}(R_{n},r)) > \mathfrak{T}(R_{n}) \vee \mathfrak{T}(R_{n}) = 1
		\tag{R2}
	\end{equation}
	
	\textmd{Binary and continuous trust metrics have been checked on R1 and R2. They failed at least on one requirement. The weighted trust metric is the first proposal by Kantert et al. \cite{Kantert2015}. The ratings are represented by float values from $[-1;1]$ (\ref{wses1}) and only on a small border case, where the per-entity rating storage is exceeded, (R2) does not hold. \\
	The second and final proposal includes the Simple Exponential Smoothing (SES) \cite{Brown1957} to form the Weighted SES Trust metric. It is an \textit{"{}advanced version of a rolling average, but it does not have to remember historic values"{}} \cite{Kantert2015}. Newer ratings have more impact. This can be adjusted via the $\alpha$ value in (\ref{wses3}). The memory usage is reduced to one value (\ref{wses1}), even though the semantic values (weights) for positive and negative ratings have to be stored independently (case distinction in (\ref{wses3})). The trust value $\tau^{s}$ is then calculated according to (\ref{wses4}).}

	\begin{equation}\label{wses1}
		\mathfrak{R}^{S} := [-1,1],r^{S} \in \mathfrak{R}^{W}
	\end{equation}
	\begin{equation}\label{wses2}
		{R}^{S} := [0,1]^{2}
	\end{equation}
	\begin{equation}\label{wses3}
		\begin{gathered}
			R^{S}_{n+1} := \mathfrak{U}^{S}(R^{S}_{n},r) \\
			:=
			\begin{cases}
				(p_{1} * \alpha + (1-\alpha) * r, p_{2} * \alpha) & r > 0, (p_{1},p_{2}) \in R^{s}_{n}\\
				(p_{1} * \alpha, p_{2} * \alpha - (1-\alpha) * r) & r < 0, (p_{1},p_{2}) \in R^{s}_{n}\\
				R^{S}_{n} & otherwise
			\end{cases}
		\end{gathered}
	\end{equation}
	\begin{equation}\label{wses4}
		\tau^{s} := \mathfrak{T}^{s} (R^{s}_{n}) := \frac{p_{1}-p_{2}}{p_{1}+p_{2}} , (p_{1},p_{2}) \in R^{s}_{n}
	\end{equation}
	
\begin{figure}[h]
		\centering
		\includegraphics[page=1, scale=0.4]{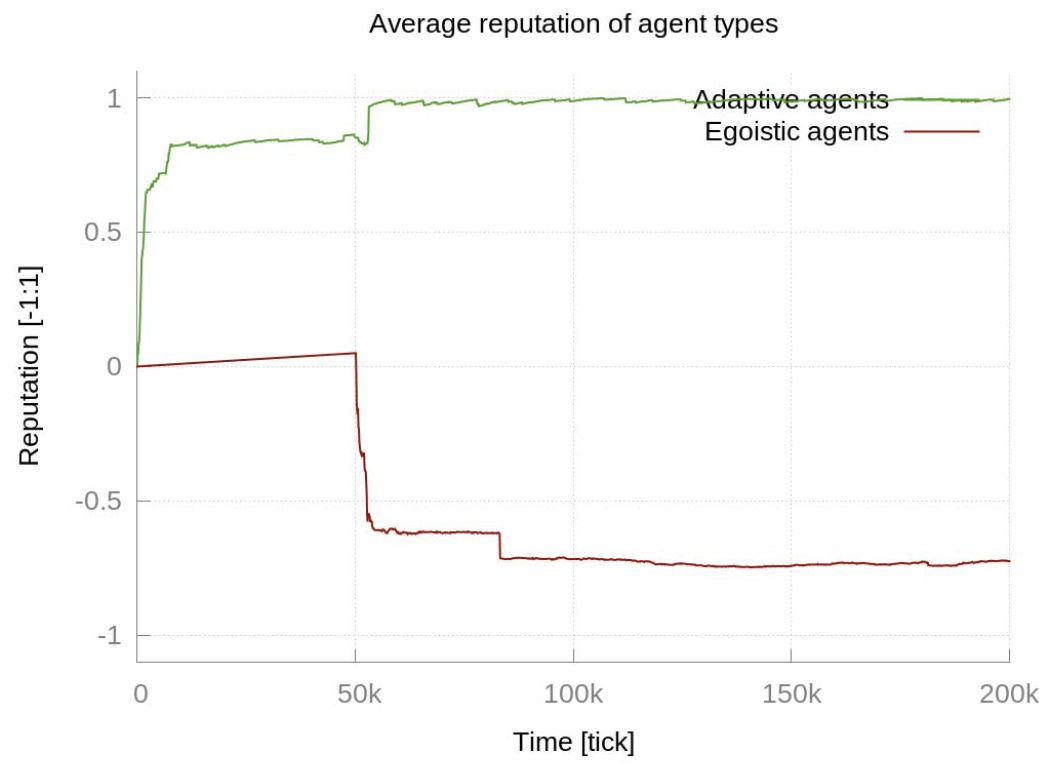}
		\caption{Reputation for metric $\tau_{s}$. 1 is the best and −1 the worst reputation. By Kantert et al \cite{Kantert2015}.}
		\label{wses_eval_plot1}
		\vspace{12pt}
		\centering
		\includegraphics[page=1, scale=0.225]{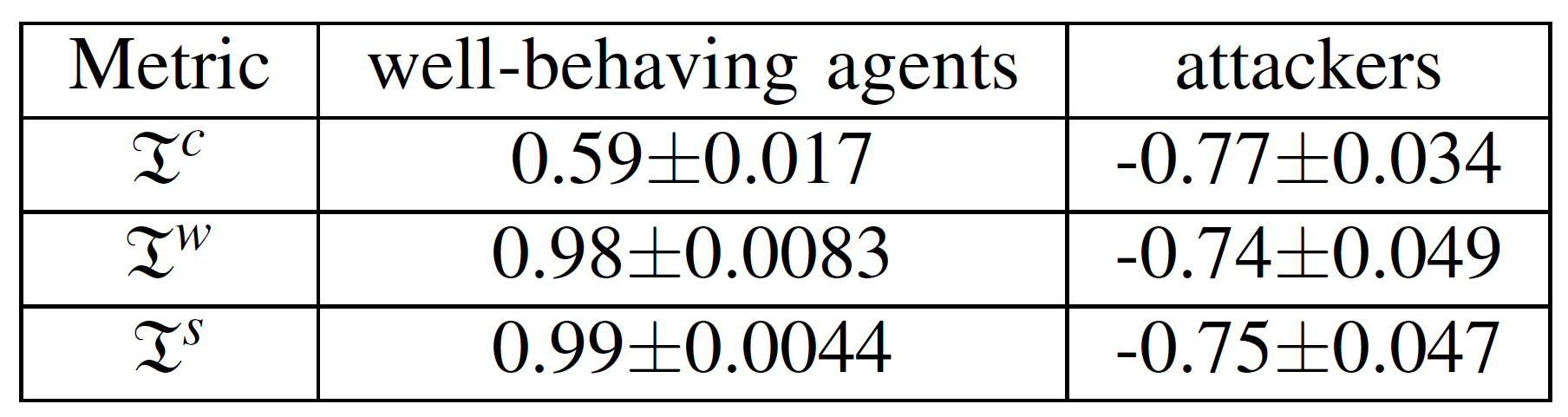}
		\caption{Reputation averaged per metric by Kantert et al \cite{Kantert2015}.}
		\label{wses_eval_plot2}
	\end{figure}

	\textmd{A proof that WSES fulfills R1 and R2 is given in \cite{Kantert2015}. To challenge the metrics on performance, a Trusted Desktop Grid (TDG) \cite{TDG2015} is set up with 100 benevolent entities which operate on tasks, when on the 50000th system tick, 100 new entities are added to the system. The new entities behave maliciously. The performance is tested in terms of speed and strength in which the malevolent entities are isolated by the initial 100 benevolent entities. Fig.~\ref{wses_eval_plot1} from Kantert et al. \cite{Kantert2015} shows the average reputation of agent types over time (ticks) when WSES trust metric was used. The attack at tick 50000 is clearly visible. While the reputation on the egoistic agents falls from $0,05$ to $-0,65$, the adaptive agents rise from $0,8$ to $1$. Fig.~\ref{wses_eval_plot2} shows a direct comparison between the continuous ($\mathfrak{T}^{c}$), weighted ($\mathfrak{T}^{w}$) and weighted simple exponential smoothed metric ($\mathfrak{T}^{s}$). The weighted and WSES metrics are significantly better in rating well-behaving entities than the continuous metric. This is reflected by the reputation averages being close to $1$. All three metrics penalize malicious behaviour by the attackers very strictly with negative reputations averaged around $-0.76$.}
	
\newpage
	
	\textmd{Because the WSES metric fulfills both requirements (R1) and (R2) unconditionally and has the overall best incentive/penalizing behaviour, it will serve as a comparison to the SimTrust metric. Differences and possibilities to combine these two approaches are served in the following section.}

\section{Comparing SimTrust and WSES}

\textmd{In the following section, SimTrust and WSES will be compared. Because of the different fundamentals of the two, it is not trivial to find a common benchmark. Therefore, both systems are compared on different use cases to point out individual advantages in the first subsection. Subsequently, concepts for combining the advantages are proposed in the second subsection.}

\begin{figure*}[b]
	\centering
	\includegraphics[width=0.6\textwidth]{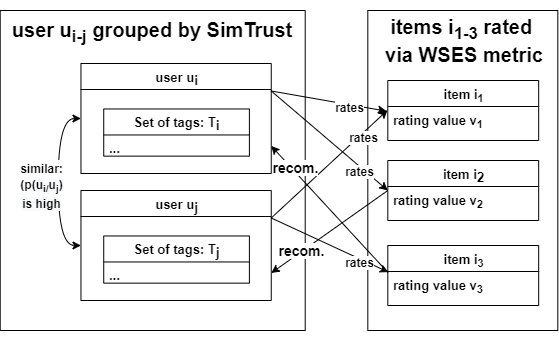}
	\caption{Approach for combining WSES and SimTrust on a database for interactable items.}
	\label{wses_simtrust_merge1}
	
	\vspace*{\floatsep}
	
	\includegraphics[width=0.6\textwidth]{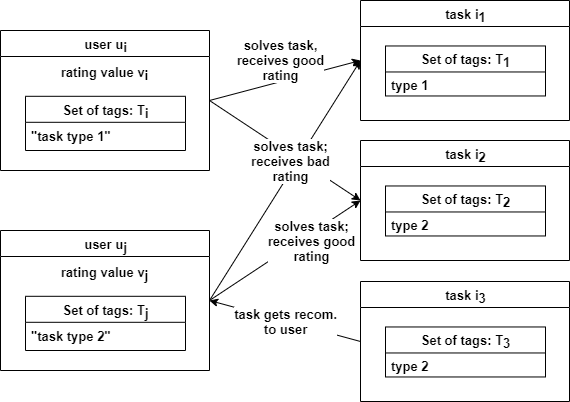}
	\caption{Approach for combining WSES and SimTrust to get additional information about entity rating.}
	\label{wses_simtrust_merge2}
\end{figure*}

\subsection{Approaches on different use cases}

	\textmd{Both of the discussed metrics aim for establishing trust among entities. While SimTrust was ultimately developed for item recommendation based on grouped entities, the main field of application for the WSES metric would be the mere isolation of malevolent entities in a SOS that follows a common goal. Finding a mutual transformation is therefore not trivial.\\
	Both metrics assume multiple entities that benefit from being grouped with entities that they trust. On SimTrust, trust among entities is derived from tagging information. The tag semantics are derived from item descriptions. Therefore, the cold-start problem is not as severe as it is on the WSES metric \cite{Simtrust2010}. So SimTrust might have advantages on freshly established systems that do not have a big database of item ratings. Another advantage of SimTrust over WSES might be on use cases where explicit rating is not common, e.g. social networks or dating platforms like Tinder \cite{Degen2020}. Here it might be more socially acceptable to group entities based on their tags than using explicit rating. Also, the overall experience for the entity might be improved when it is not only grouped with other entities based on geographical properties but also on common features. \\
	WSES does not hold any items. Entities fulfil tasks to gain an explicit rating. In comparison to SimTrust, this might be more accurate and reasonable for entities of these systems. They get fair ratings, ensured by R1 and R2. SimTrust might be incompatible for evaluation on R1 and R2 due to the missing of explicit trust values. Trust from tagging mostly relies on numeric values for calculating the similarity of tags, not for an entity's rating. So in SimTrust are neither explicit ratings used nor is any comparable metric involved.\\
	On WSES, entities that do not follow the common system goal will get strictly isolated from the rest of the system. In this case isolation means that no more tasks are delegated to the entity but he still might be able to submit new work units. He will be inactive as a working unit though. On SimTrust this would result in not receiving any recommendations. When comparing to the strength of penalizing on WSES, this penalty on SimTrust would be less severe because an entity can still be active in the SimTrust environment even though it does not get any recommendations.\\
	After outlining advantages and disadvantages of the systems over each other, combining them for better overall results will be approached in the next subsection.}

\subsection{Different combinations of WSES and SimTrust}

	\textmd{A first idea of merging the two approaches might be using SimTrust as a base system for trust among entities but with a modified ranking for items in a database via the WSES. This is presented in Fig.~\ref{wses_simtrust_merge1}. First, entities are grouped via their tags. When they have interacted with items, they can rate them via the WSES metric. For example should an item be recommended to an entity A when similar entities have already interacted with the item and gave it a good rating. This would help distinguishing entities with their different requirements on an item. There is for example an e-commerce system, where entities can purchase items. Entity A characterizes itself with the tags ${quality, haptic, material}$. It rates a football P with 2 out of 5 possible points. Entity B characterizes itself with the tags ${look, beauty, appearance}$. It rates P with 5 out of 5 possible points. The item will now only be recommended to entities that are similar to B because they might set the same requirements and be as satisfied as B is.}

	\textmd{Entity-dependant relevance of rating can also be achieved differently. Enriching the numeric rating with a set of tags will precise the meaning of a rating. From this, multiple use cases are possible: For example, entities that share tags with the tagged ratings can only view these or items with sufficient ratings and shared tags get recommended only to fitting entities.}


\textmd{A second idea on combining SimTrust with WSES could be used on use cases similar to TDG. Even though WSES has a fair metric for trust presented in one value, it does not hold any more information for interpretation. This could be inefficient when there are entities that have different abilities. Just from a single trust value, it is not possible to differentiate the special skills of the individual entities when they have similar trust values. They would only be considered from "{}good"{} to "{}bad"{}. Tags could enrich the information about an entity, describing its abilities. Fig.~\ref{wses_simtrust_merge2} visualizes the idea.}

\textmd{A possible use case could be a SOS with the goal of solving mathematical equations. They may include integrations or matrix multiplication. The entities in the SOS have individual skills in solving the equations. User $u_{i}$ is able to solve integrations, while user $u_{j}$ is only able to solve matrix multiplications. When $u_{i}$ solves the task $i_{1}$ which is an integration task, it will receive a good rating because it is specialized for this type of task and can perform it in a satisfactory way. It will not receive a good rating on task $i_{2}$ because it does not have the skills to solve it. User $u_{j}$ performs well on $i_{2}$ but not good on $i_{1}$. The task type where the user performed well will be saved in their tags so it will be matched or selects itself further tasks ($i_{3}$) where it is likely to receive good ratings. This behaviour encourages specialization of users, even more if the users are able to improve their performance on runtime. To establish a credibility of ratings and abilities, a network of trust could be set up based on the transitivity of trust. In other words, entities can also mediate between others that do not share an interaction history.}


\section{Conclusion}


\textmd{SimTrust and WSES have different ways of approaching trust among entities. While WSES relies on explicit rating from thirds, SimTrust is based on entity tagging. This causes WSES to perform well on systems that encourage explicit rating, while SimTrust has advantages on fresh systems because the cold-start problem can be reduced. It is difficult to compare the systems on their performance because of these different implementations. Even though both aim for trust, the usage later on in the applications is different so the requirements on the trust metrics vary heavily. WSES is mainly used on collaborative problem solving while SimTrust has been developed for item recommendation. However, two ideas for combining the metrics could be stated. Both ideas use the tagging information to enrich explicit ratings. The first concept aims on recommending items with good ratings to entities that are trusting the reviewers. The other concept recommends tasks to users which are likely to perform well on a task derived from the required abilities.}

\textmd{The initial question, which of the presented systems should be used if trust has to be implemented on a typical SOS, has evolved into a comparison of SimTrust and WSES. The different nature of these two has been described and it could be seen that different advantages and disadvantages are arising from this. Therefore it was not possible to find a common use case on which they could have been ranked in terms of performance. Both systems have their advantages depending on the detailed implementation of the respective SOS. It is not realistic to compare SimTrust and WSES on a so-called "{}default"{} SOS. Every open SOS already brings features that puts one of the metrics in favour. However, elaborating the advantages on different use cases exposed benefits which were able to be combined into new metric concepts. In the future, it would be interesting to evaluate these concepts in a real scenario as well as researching more on a common scenario for SimTrust and WSES to get an explicit comparison between these two.}


\bibliographystyle{IEEEtran}

\bibliography{mybib}
\end{document}